\documentclass[aps,prl,showpacs,twocolumn]{revtex4}
\usepackage{amsfonts}
\usepackage{amssymb}
\usepackage{amsmath}
\usepackage{graphicx}
\usepackage{epsfig}

\input{tcilatex}

\begin{document}

\title{Coherent shift of localized bound pair in Bose Hubbard model}
\author{L. Jin, B. Chen, and Z. Song$^{\dag}$ }
\affiliation{Department of Physics, Nankai University, Tianjin 300071, China}

\begin{abstract}
Based on the exact results obtained by Bethe ansatz, we demonstrate the
existence of stable bound pair (BP) wave packet in Bose Hubbard model with
arbitrary on-site interaction $U$. In large-$U$ regime, it is found that an
incoming single-particle (SP) can coherently pass through a BP wave packet
and leave a coherent shift in the position of it. This suggests a simple
scheme for constructing a BP charge qubit to realize a quantum switch, which
is capable of controlling the coherent transport of one and \textit{only one}
photon in a one-dimensional waveguide.
\end{abstract}

\maketitle

\textit{Introduction.} Most recently, many theoretical and experimental
investigations about bound pair (BP) in strongly correlated boson systems
are carried out \cite{Winkler, Mahajan, Petrosyan, Creffield, Kuklov,
Folling, Zollner, ChenS} since the experimental observation of atomic BP in
optical lattice \cite{Winkler}. Counter intuitively, it is found that the
trapped rubidium atoms in a three-dimensional optical lattice can form a
stable BP, even though in free space the two atoms would have repelled each
other. For the problems BP, we can cast back for much earlier investigations
of $\eta $-pairing states in Hubbard model for electrons, which possess
off-diagonal long-range order (ODLRO) \cite{YangCN}. Actually, the basic
physics of both fermion and boson BPs is that, the periodic potential
suppresses the single particle tunnelling across the barrier, a process that
would lead to a decay of the pair. Interesting questions are whether such a
BP as composite particle will occur in moderate $U$ system and whether it
can exist stably as a wave packet. It is crucial for quantum information
processing since the Bose Hubbard model is the simplest model capturing the
main physics of not only cold atoms in optical lattice also photons in
nonlinear waveguide \cite{Jaksch,QBEC,cavities}.

In this paper we will present some exact results obtained by Bethe ansatz
concerning the two-particle problem. We demonstrate the existence of a
stable BP wave packet in Bose Hubbard model with arbitrary on-site
interaction $U$. It is found that the most statble BP wave packets refer to
different regions of a bound pair band (BPB) and have different group
velocities as $U$ varies from zero to infinity, but spread to the same
fidelity when they travel over the same distance. This feature allows the BP
wave packet as a new object to be a flying and stationary qubit in quantum
device. We also investigate the scattering between a BP wave packet and a
single particle (SP) in large-$U$ limit. It is found that an incoming SP
wave packet can coherently pass through a BP and leave a coherent shift in
the position of the BP, which arises from the exotic effective exchange
interaction between them. Furthermore, utilizing on-site $U$ one can confine
a BP, rather than a SP, in two sites to form a charge qubit. This suggests a
simple scheme to realize a quantum switch, which is capable of controlling
the coherent transport of one and \textit{only one} photon in a
one-dimensional waveguide.

\textit{Wave packet in bound-pair band. }The simplest model capturing some
physics of the nonlinearity of photons in a coupled cavity array and cold
atoms in optical lattice is a Bose Hubbard model. The Hamiltonian $H$ is
written as follows: 
\begin{equation}
H=-\kappa \overset{N}{\sum_{i=1}}\left( a_{i}^{\dag }a_{i+1}+h.c.\right) -%
\frac{U}{2}\overset{N}{\sum_{i=1}}n_{i}(n_{i}-1),  \label{H}
\end{equation}%
where $a_{i}^{\dag }$ is the creation operator of the boson at $i$th site,
the tunnelling strength and on-site interaction between bosons are denoted
by $\kappa $\ and $U$. For the sake of clarity and simplicity, we only
consider odd-site system with $N=2N_{0}+1$, and periodic boundary conditions%
\textbf{?} $a_{N+1}=a_{1}$.

Consider the two-particle problem, a state in the two-particle Hilbert space
can be written as \label{basis} 
\begin{subequations}
\begin{eqnarray}
\left\vert \psi _{k}\right\rangle &=&\sum_{k,r}f^{k}(r)\left\vert \phi
_{r}^{k}\right\rangle , \\
\left\vert \phi _{0}^{k}\right\rangle &=&\frac{1}{\sqrt{2N}}e^{i\frac{k}{2}%
}\sum_{j}e^{ikj}\left( a_{j}^{\dag }\right) ^{2}\left\vert Vac\right\rangle ,
\\
\left\vert \phi _{r}^{k}\right\rangle &=&\frac{1}{\sqrt{N}}e^{i\frac{k(r+1)}{%
2}}\sum_{j}e^{ikj}a_{j}^{\dag }a_{j+r}^{\dag }\left\vert Vac\right\rangle ,
\end{eqnarray}%
where $k=2\pi n/N$, $n\in \lbrack 1,N]$ denotes the momentum, and $r\in
\lbrack 1,N_{0}-1]$ is the distance between two particles. Due to the
translational symmetry of the present system, the Schr$\ddot{o}$dinger
equations for $f^{k}(r)$, $r\in \lbrack 0,N_{0}-1]$ is easily shown to be 
\end{subequations}
\begin{equation}
\lbrack \sum_{j=0}^{N_{0}-1}T_{j}^{k}\left( \delta _{j,r+1}+\delta
_{j,r-1}\right) -U\delta _{r,0}+T_{r}^{k}\delta _{r,N_{0}}-\varepsilon
_{k}]f^{k}(r)=0,  \label{Schrodinger}
\end{equation}%
where $T_{r}^{k}=-2\sqrt{2}\kappa \cos (k/2)$ for $r=0$, and $-2\kappa \cos
(k/2)$ for $r\neq 0$, respectively. Obviously, for an arbitrary $k$, the
solution of (\ref{Schrodinger}) is equivalent to that of a non-interacting $%
N_{0}$-site tight-binding chain with nearest neighbor hopping amplitude $%
T_{j}^{k}$,\ on-site potentials $U$ and $-2\kappa \cos (k/2)$\ at two ends
respectively. In this work, we focus our study on the bound states. In each $%
k$-invariant subspace, there exists only one bound state for nonzero $U$,
which can be obtained via Bethe ansatz method. And all the $N$ bound states,
indexed by $k$,\ constitute a bound-pair band (BPB). 
%%%%%%%%%%%%%%%%%%%%%%%%%%%%%%%%%%%%%%%%%%%%%%%%%%%%%%%%%%%%%%%%%
\begin{figure}[tbp]
\includegraphics[ bb=11 159 582 780, width=3.9 cm, clip]{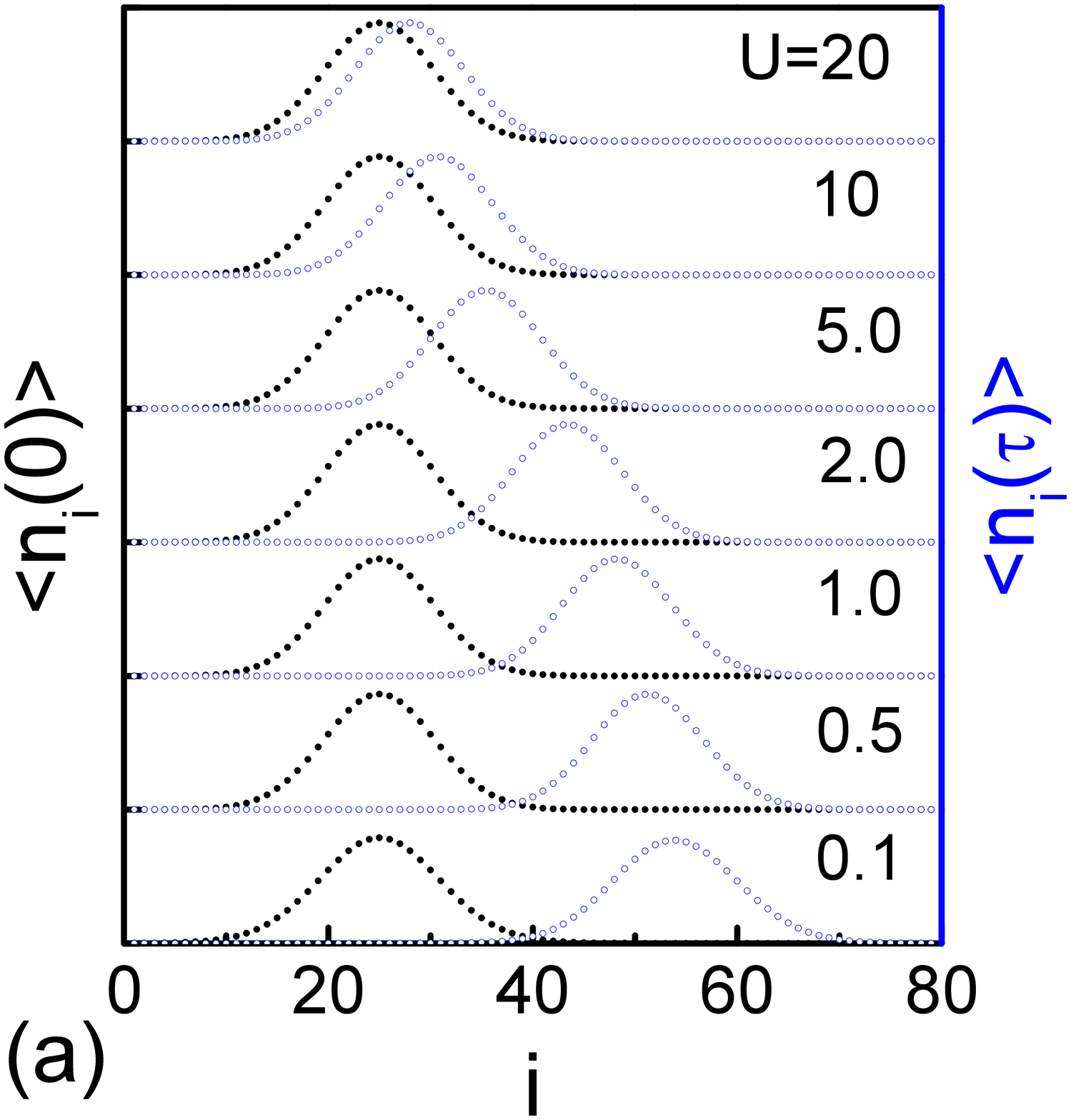} %
\includegraphics[ bb=19 229 582 770, width=4.4 cm, clip]{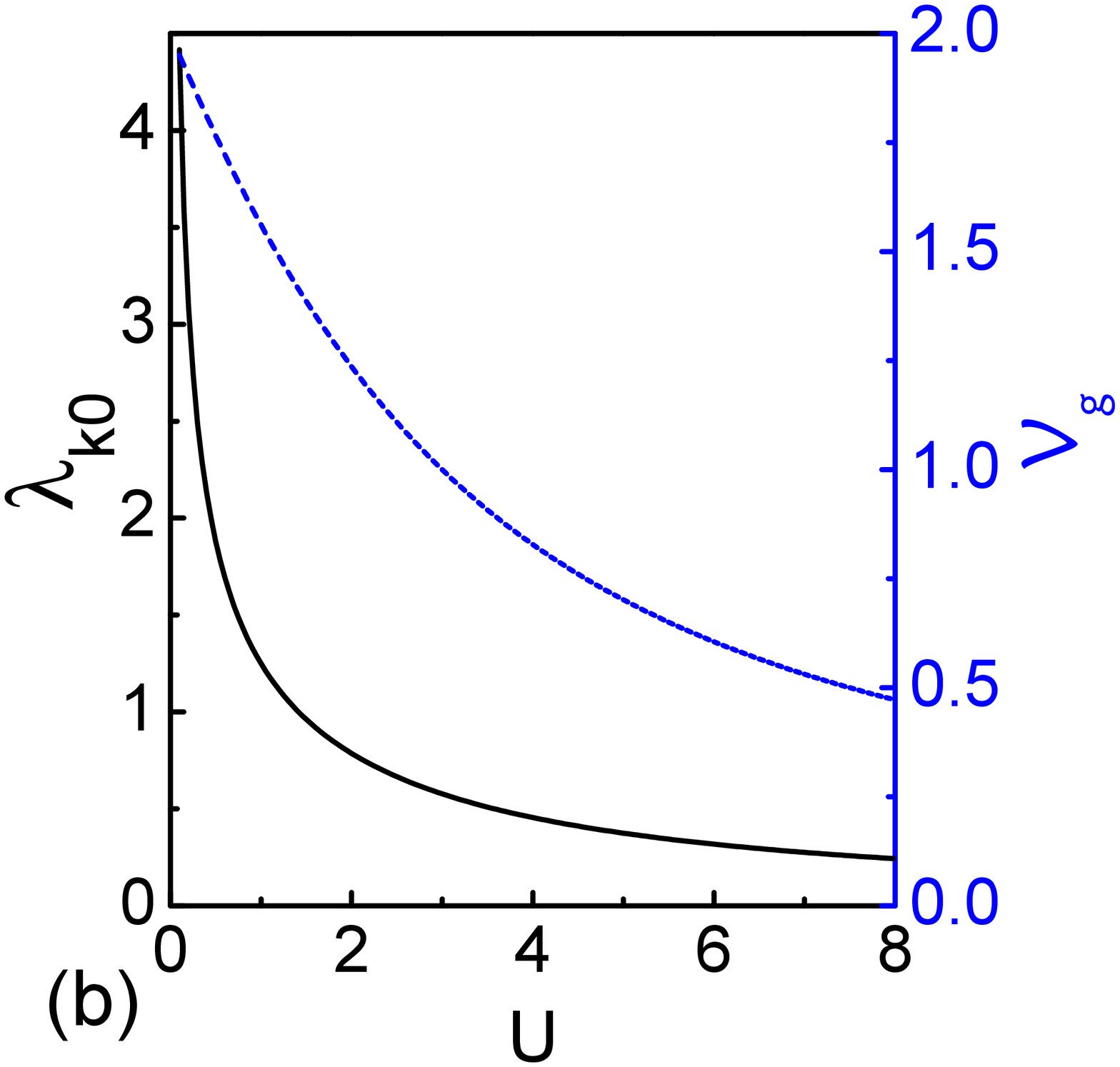} %
\includegraphics[ bb=104 340 502 484, width=7.0 cm, clip]{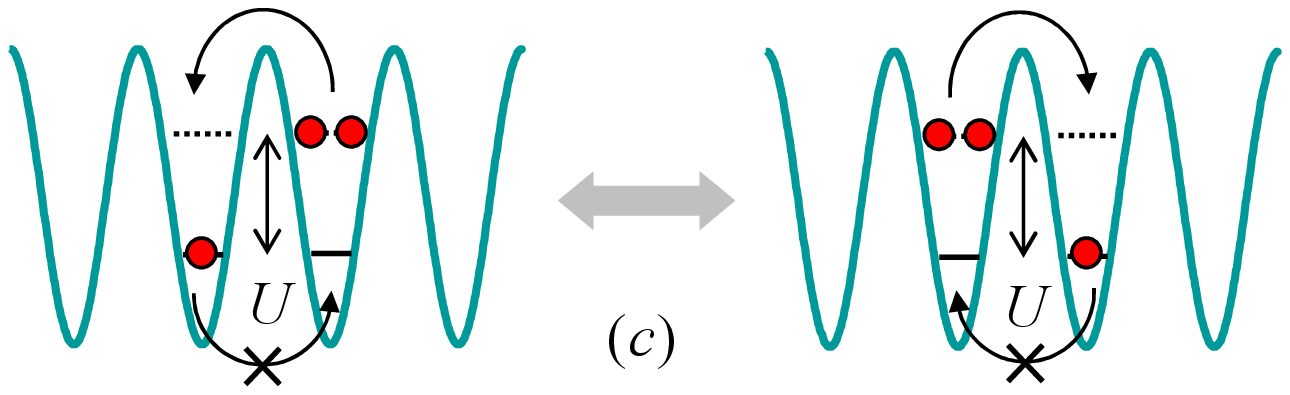}
\caption{\textit{(Color online) (a) Plots of }$\left\langle
n_{i}(t)\right\rangle $\textit{\ for the wave packets with }$\protect\alpha %
=2/15$\textit{\ at time }$t=0$ \textit{(dot)} and $t=\protect\tau =15/%
\protect\kappa $\textit{\ (empty circle) in the systems with }$U=0.1\mathit{%
\sim }20$\textit{\ in the unit of }$\protect\kappa $\textit{. (b) Plots of }$%
v_{g}$ \textit{in the unit of }$\protect\kappa $\textit{\ (dashed line) and
size }$\protect\lambda _{k_{0}}$\textit{\ (solid line) for the BP Gaussian
wave packets. (c) Schematic illustration of the exchange\ interaction
between SP and BP.}}
\label{fig1}
\end{figure}
%%%%%%%%%%%%%%%%%%%%%%%%%%%%%%%%%%%%%%%%%%%%%%%%%%%%%%%%%%%%%%%%%

For a large $N$ system, the pair-bound band can be expressed as 
\begin{equation}
\varepsilon _{k}=\text{\textrm{sgn}}(U)\sqrt{U^{2}+16\kappa ^{2}\cos ^{2}%
\frac{k}{2}}  \label{e(k)}
\end{equation}%
with the wave function%
\begin{equation}
f^{k}(r)\simeq \lbrack \text{\textrm{sgn}}(\zeta _{k})]^{r}\left( 1+\zeta
_{k}^{2}\right) ^{-\frac{1}{4}}\left\{ 
\begin{array}{c}
1\text{, \ \ \ \ \ \ }(r=0) \\ 
\sqrt{2}e^{-\left\vert \mu _{k}\right\vert r}\text{,\ }(r\neq 0)%
\end{array}%
\right. ,  \label{f(r)}
\end{equation}%
where $\zeta _{k}=4\kappa \cos (k/2)/U$ and $\mu _{k}=\ln [1/\zeta _{k}+%
\sqrt{1+\left( 1/\zeta _{k}\right) ^{2}}]$. The spectrum of BP (\ref{e(k)})
is in agreement with that obtained from the Green's function method \cite%
{Winkler}. The size of the BP for every bound state can be characterized by 
\begin{equation}
\lambda _{k}=\sqrt{\sum_{r}\left\vert rf^{k}(r)\right\vert ^{2}}=\left\vert 
\frac{2\sqrt{2}\kappa }{U}\cos \left( \frac{k}{2}\right) \right\vert ,
\label{lambda(k)}
\end{equation}%
which depends not only on $\kappa /U$\ but also on $k$. It can be seen that
even for weak $U$, the size of BP\ still\ remains small\ for long-wave
eigenstates, which is crucial for the following discussion. On the other
hand, for each eigenstate, the BP is delocalized as a composite particle.
Nevertheless, it has been argued that approximately nonspreading wave packet
can be achieved by a superposition of eigenstates within a linear region, so
as to the populated energy levels are equally spaced \cite{LY,YS}. Note that
there exists a linear region in the vicinity of $k_{0}$ in the BPB\ spectrum
(\ref{e(k)}) for any value of $U$. Here $k_{0}$\ is determined by the
condition $\left( \partial ^{2}\varepsilon _{k}/\partial k^{2}\right)
_{k=k_{0}}=0$\ or its more explicit form $\cos k_{0}=$ $\sqrt{\eta ^{2}-1}%
-\eta $, where $\eta =$ $\left( U^{2}/8\kappa ^{2}+1\right) $. Within such a
region, a Gaussian wave packet can be constructed in the form%
\begin{equation}
\left\vert \Phi (k_{0},N_{c})\right\rangle =\frac{1}{\sqrt{\Omega }}%
\sum_{k}e^{-\frac{1}{2\alpha ^{2}}\left( k-k_{0}\right) ^{2}-iN_{c}\left(
k-k_{0}\right) }\left\vert \psi _{k}\right\rangle ,  \label{GWP(k)}
\end{equation}%
where $N_{c}\in \lbrack 1,N]$ is the center of it in real space, and \ $%
\Omega =\sum\nolimits_{k}e^{-\frac{1}{\alpha ^{2}}\left( k-k_{0}\right) ^{2}}
$ is the normalization factor. The dynamics of such a wave packet\ is
governed by the effective Hamiltonian $H_{eff}=$ $\sum_{k}\tilde{\varepsilon}%
_{k}\left\vert \psi _{k}\right\rangle \left\langle \psi _{k}\right\vert $
approximately. Here the effective linear dispersion relation is $\tilde{%
\varepsilon}_{k}=\varepsilon _{k_{0}}+v_{g}\left( k-k_{0}\right) $, where 
\begin{equation}
v_{g}=\left( \frac{\partial \varepsilon _{k}}{\partial k}\right)
_{k=k_{0}}=2\kappa \sqrt{1-\frac{U^{2}}{8\kappa ^{2}}\left( \sqrt{1+\frac{%
16\kappa ^{2}}{U^{2}}}-1\right) }  \label{v_g}
\end{equation}%
is the group velocity of the wave packet (\ref{GWP(k)}) in real space.

Now we firstly investigate the dynamics of such wave packet for any value of 
$U$ under the linear approximation. Taking the state (\ref{GWP(k)})\ as an
initial state $\left\vert \tilde{\Psi}(t=0)\right\rangle $, its time
evolution driven by $H_{eff}$\ presents $\left\vert \tilde{\Psi}%
(t)\right\rangle =$ $e^{-iH_{eff}t}\left\vert \Phi
(k_{0},N_{c})\right\rangle $ $=e^{i\varphi }\left\vert \Phi
(k_{0},N_{c}+v_{g}t)\right\rangle $. The overall phase factor $e^{i\varphi }$
has no effect on the final result. It is obvious that the wave packet moves
along the ring with velocity $v_{g}$. In this sense, the time evolution of
some states governed by $H_{eff}$ can be described as a spatial translation
by the operator $\mathcal{U}(t)=\exp (-ikv_{g}t)\equiv \mathcal{T}(l)$ with
a displacement $l=v_{g}t$. This shows that the shape of the wave packet in
the real space does not change approximately during its travel. However, for
the exact time evolution, state $\left\vert \Psi (t)\right\rangle =$ $%
e^{-iHt}\left\vert \Phi (k_{0},N_{c})\right\rangle $ is slightly different
from state $\left\vert \tilde{\Psi}(t)\right\rangle $ due to the
nonlinearity of the dispersion (\ref{e(k)}). The overlap between two states
that evolve from the same initial wave function under two different
Hamiltonians $H$\ and $H_{eff}$, respectively,\ is defined as the Loschmidt
echo (LE) or quantum fidelity $F(t)=\langle \Psi (t)\left\vert \tilde{\Psi}%
(t)\right\rangle $ which can be employed to depict the deformation of a
travelling wave packet. A straightforward calculation shows that%
\begin{equation}
F(t)=\frac{1}{\Omega }\sum_{k}e^{-\left( k-k_{0}\right) ^{2}/\alpha
^{2}}\cos \left[ \frac{1}{6}\left( k-k_{0}\right) ^{3}v_{g}t\right] ,
\label{F(t)1}
\end{equation}%
which is based on the fact $\left( \partial ^{3}\varepsilon _{k}/\partial
k^{3}\right) _{k=k_{0}}=-v_{g}$.

Remarkably, the fact that the fidelity (\ref{F(t)1}) only depends on $v_{g}t$
means that the wave packets with fixed $\alpha $\ but different $k_{0}$
share the same fidelities after they travel the same distance $l=v_{g}t$. It
indicates that a slower wave packet in strong $U$ system has longer life
time comparing to a faster one in a weak $U$ system. This feature can be
utilized to quantum information processing: weak $U$ system can be a quantum
channel for quantum state transfer, while strong $U$ system can be employed
for quantum state storage.

The above discussion tells us that a state of type (\ref{GWP(k)}) is
non-spreading only within certain approximate limits. Next we investigate
the profile of such a state in real space. It is well known that for a SP
case, if we replace $\left\vert \psi _{k}\right\rangle $ as $\left\vert
k\right\rangle =1/\sqrt{N}\sum_{j}e^{ikj}a_{j}^{\dag }\left\vert
Vac\right\rangle $, the SP wave function is%
\begin{equation}
\left\vert \phi (k_{0},N_{c})\right\rangle =\frac{1}{\sqrt{\Omega }}%
\sum_{j}e^{-\frac{^{\alpha ^{2}}}{2}(j-N_{c})^{2}+ik_{0}j}a_{j}^{\dag
}\left\vert Vac\right\rangle ,  \label{single P}
\end{equation}%
which is also a wave packet of Gaussian type \cite{YS}. For two-particle
state (\ref{GWP(k)}), its profile in real space can be described by the
distribution of the average particle density%
\begin{equation}
\left\langle n_{i}(t)\right\rangle =\left\langle \Psi (t)\right\vert
a_{i}^{\dag }a_{i}\left\vert \Psi (t)\right\rangle =\left\vert
a_{i}\left\vert \Psi (t)\right\rangle \right\vert ^{2}.  \label{n_l(t)}
\end{equation}%
In Fig. 1 (a) we plot $\left\langle n_{i}(t)\right\rangle $\ for the wave
packets with $\alpha =2/15$\ at time $t=0$ and $\tau =15/\kappa $ in the
systems with $U=0.1$, $0.5$, $1$, $5$, $10$, and $20$ in the unit of $\kappa 
$. It shows that the profile of the wave packets in the real space is also
Gaussian type and non-spreading. It also indicates that the shape of the
wave packets does not change apparently for different $U$, which is in
agreement with the following observation from Eq. (\ref{Lambda_k0}) that the
size of a BP remains small for $\left\vert U\right\vert \succsim 0.2\kappa $.

For a given $\alpha $, the size of a composite particle in the form of a
nonspreading wave packet\ is a function only of the ratio $\kappa /U$

\begin{equation}
\lambda _{k_{0}}=\frac{1}{\sqrt{2}}\sqrt{\sqrt{1+\frac{16\kappa ^{2}}{U^{2}}}%
-1},  \label{Lambda_k0}
\end{equation}%
which determines the size of the wave packet. To demonstrate the features of
a BP wave packet for arbitrary $U$ system, its velocity $v_{g}$\ and size $%
\lambda _{k_{0}}$ are plotted in Fig. 1 (b). For $U/\kappa =0.2\sim 10$, we
have $\lambda _{k_{0}}=0.2\sim 3.1,$ which is sufficiently small that we can
have many pairs in the lattice without having substantial overlap between
them. It implies a new phase, a gas of BPs which has been predicted in large 
$U$ limits \cite{Petrosyan}, can also exist in moderate $U$ system. It is
worthy to stress that, although a SP wave packet (\ref{single P}) and a BP
wave packet (\ref{GWP(k)}) share some common properties, on-site interaction 
$U$ is able to govern a BP rather than a SP wave packet. In this sense, the
SP and \qquad BP wave packets can be regarded as two different types of
particles. Nevertheless the interaction between them is exotic since the
constituent of BP is essentially SP.

\textit{Coherent shift. }In the following we restrict ourselves to large $U$%
\ limit. Consider a three-body problem. The spectrum consists of three bands
around $0$, $U$ and $2U$. We are interested in the middle band, which
corresponds to a SP and a BP. Using perturbation method, the corresponding
effective Hamiltonian is%
\begin{eqnarray}
\tilde{H} &=&-\kappa \sum_{i=1}^{N}\tilde{a}_{i}^{\dag }\tilde{a}_{i+1}+%
\frac{4\kappa ^{2}}{U}\sum_{i=1}^{N}\tilde{b}_{i}^{\dag }\tilde{b}_{i+1}
\label{H hardcore} \\
&&-\sqrt{2}\kappa \sum_{i=1}^{N}\tilde{b}_{i+1}^{\dag }\tilde{b}_{i}\tilde{a}%
_{i}^{\dag }\tilde{a}_{i+1}+h.c.+U\sum_{i=1}^{N}\tilde{b}_{i}^{\dag }\tilde{b%
}_{i},  \notag
\end{eqnarray}%
where $\tilde{a}_{i}$\ and $\tilde{b}_{i}$\ denote the hardcore bosons
satisfying the following commutation relations $[\tilde{a}_{j},\tilde{a}%
_{i}^{\dag }]=$ $[\tilde{b}_{j},\tilde{a}_{i}^{\dag }]=$ $[\tilde{b}_{j},%
\tilde{b}_{i}^{\dag }]=0,$ $(i\neq j)$; $\{\tilde{a}_{i},\tilde{a}_{i}^{\dag
}\}=$ $\{\tilde{b}_{i},\tilde{a}_{i}^{\dag }\}=$ $\{\tilde{b}_{i},\tilde{b}%
_{i}^{\dag }\}=0$. 
%%%%%%%%%%%%%%%%%%%%%%%%%%%%%%%%%%%%%%%%%%%%%%%%%%%%%%%%%%%%%%%%%%%%%%%%
\begin{figure}[tbp]
\includegraphics[ bb=67 265 555 741, width=6.0 cm, clip]{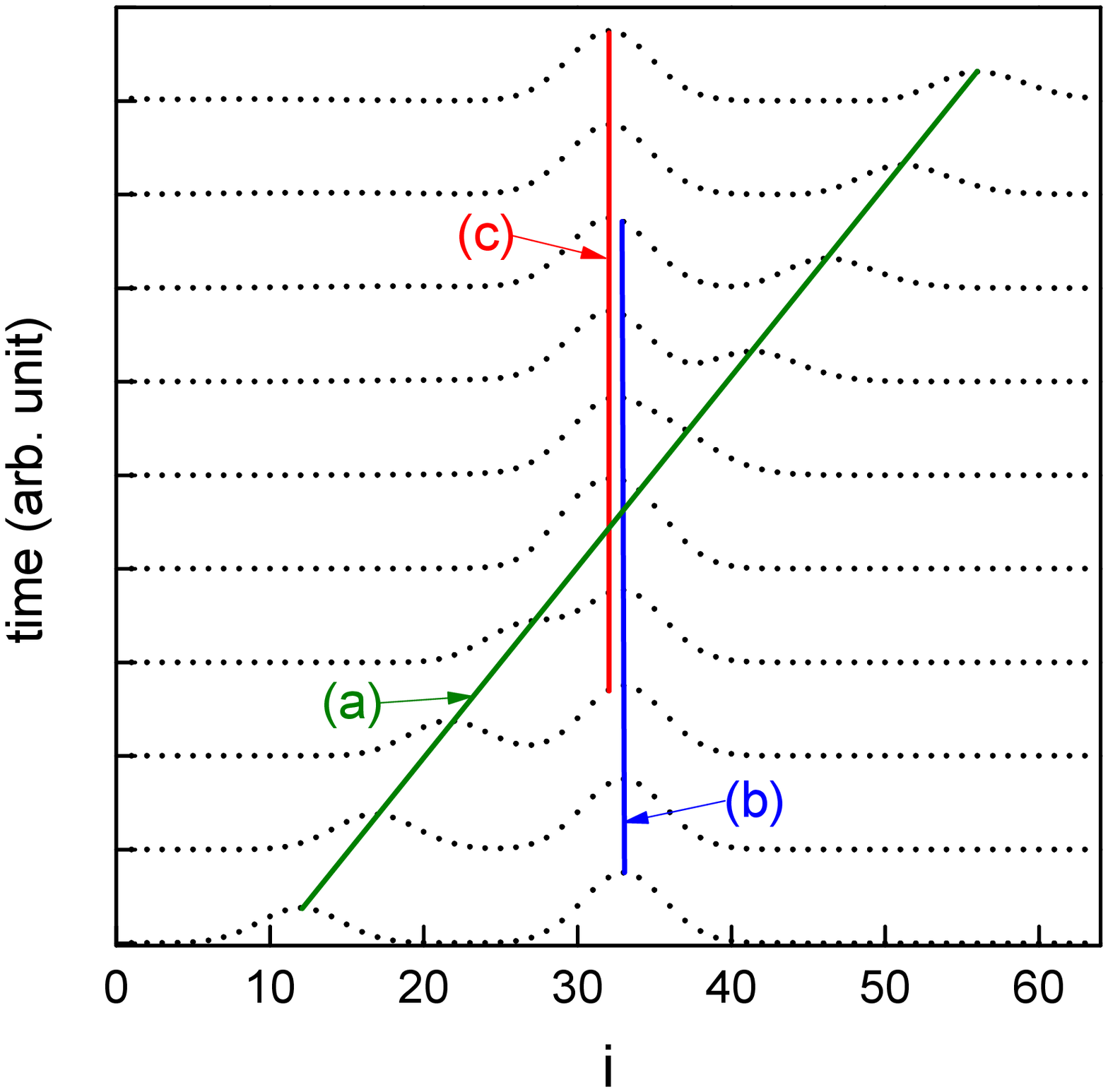}
\caption{\textit{(Color online) The stroboscopic picture\ of the profiles of
evolving SP and BP wave packets obtained by numerical simulations: line }$a$%
\textit{\ denotes the center of SP wave packet, while lines }$b$\textit{\
and }$c$\textit{\ denote the centers of BP wave packets before and after
scattering. }}
\label{fig2}
\end{figure}
%%%%%%%%%%%%%%%%%%%%%%%%%%%%%%%%%%%%%%%%%%%%%%%%%%%%%%%%%%%%%%%%%%%%%%%%%
The first two terms describe the hopping of SP and BP, the third term
describe the interaction between the two kinds of particles, the process of
which is schematically illustrated in Fig. 1 (c). Now we focus on the
scattering between SP and BP wave packets. In short time duration, a BP is
relative stationary comparing with a moving SP wave packet. Then the
swapping operation $\tilde{b}_{i+1}^{\dag }\tilde{b}_{i}\tilde{a}_{i}^{\dag }%
\tilde{a}_{i+1}$\ allows the incoming SP wave packet \textquotedblleft pass
through\textquotedblright\ the BP and shift its position with a unit lattice
spacing. To demonstrate this process, numerical simulation is performed for
the time evolution of such two wave packets. The initial state is $%
\left\vert \phi (\pi /2,N_{sp})\right\rangle \left\vert \Phi (\pi
/2,N_{bp})\right\rangle $ with $N_{sp}\ll N_{bp}$. In the simplest case of
replacing the swapping term $\sqrt{2}\kappa $ by $\kappa $, the final state
is $\left\vert \phi (\pi /2,N_{sp})\right\rangle \left\vert \Phi (\pi
/2,N_{bp}-1)\right\rangle $ with $N_{sp}\gg N_{bp}$.\ Actually, factor $%
\sqrt{2}$ can cause a slight reflection of the incoming SP wave packet from
the above fact.\ Fig. 2 is the stroboscopic picture\ of the profiles of two
evolving wave packets obtained by numerical simulations for the Hamiltonian (%
\ref{H hardcore}): line $a$ denotes the center of SP wave packet, while
lines $b$ and $c$ denote the centers of BP wave packets before and after
scattering. It is clear that the incoming SP wave packet keeps the same
speed during the whole process, while the BP wave packet get a coherent
shift with a unit of lattice spacing.

\textit{Bound-pair charge qubit. }Now we apply the novel feature of coherent
shift of a BP to construct a quantum switch. Considering a $4$-site chain in
strong on-site interaction limit, the dynamics of a SP and a BP obeys the
Hamiltonian

\begin{eqnarray}
H_{CQ} &=&-\kappa \left( \tilde{a}_{s}^{\dag }\tilde{a}_{s+1}+\tilde{a}%
_{s+2}^{\dag }\tilde{a}_{s+3}\right) -\kappa _{0}\tilde{a}_{s+1}^{\dag }%
\tilde{a}_{s+2}  \label{H_cq} \\
&&+\frac{4\kappa _{0}^{2}}{U_{0}}\tilde{b}_{s+1}^{\dag }\tilde{b}_{s+2}-%
\sqrt{2}\kappa _{0}\tilde{b}_{s+1}^{\dag }\tilde{b}_{s+2}\tilde{a}%
_{s+2}^{\dag }\tilde{a}_{s+1}  \notag \\
&&+h.c.+U\sum_{i=s,s+3}\tilde{b}_{i}^{\dag }\tilde{b}_{i}+U_{0}%
\sum_{i=s+1,s+2}\tilde{b}_{i}^{\dag }\tilde{b}_{i},  \notag
\end{eqnarray}%
which is schematically illustrated in Fig. 3 (a). We focus on the case of
that there is a single BP in the site $s+1$\ and $s+2$, which can be
realized under the condition $U\gg U_{0}$. Notice that this setup is
equivalent to the system of confining a composite in an effective
double-well potential and can be regarded as a charge qubit. Such a qubit
has a novel feature due to the coherent shift induced by the scattering with
a SP. To demonstrate this, we take $\sqrt{2}\kappa _{0}=\kappa $ for
simplicity and study the dynamical process via time evolution. We embed such
a charge qubit into a chain as illustrated in Fig. 3 (b, c). Let us firstly
assume that initially the qubit is in the \textquotedblleft
right\textquotedblright\ state $\left\vert R\right\rangle =\tilde{b}%
_{s+2}^{\dag }\left\vert Vac\right\rangle $, while a SP wave packet of type (%
\ref{single P}) $\left\vert \phi (\pi /2,N_{c}\prec s)\right\rangle \equiv
\left\vert \phi (\pi /2,L)\right\rangle $ (similarly, we define $\left\vert
\phi (\pm \pi /2,N_{c}\succ s+2)\right\rangle \equiv \left\vert \phi (\pm
\pi /2,R)\right\rangle $) is coming from the left. Comparing to the speed of
the SP wave packet $v_{g}=2\kappa $, state $\left\vert R\right\rangle $\ can
be regarded as a stationary state during the whole scattering process. Then
according to the Hamilltonian $H_{CQ}$, the incoming wave will pass through
the qubit freely but leaves the qubit to be in the \textquotedblleft
left\textquotedblright\ state $\left\vert L\right\rangle =\tilde{b}%
_{s+1}^{\dag }\left\vert Vac\right\rangle $, i.e., $\left\vert \phi (\pi
/2,L)\right\rangle \left\vert R\right\rangle \longrightarrow \left\vert \phi
(\pi /2,L)\right\rangle \left\vert L\right\rangle $. In contrast, if the
qubit is in state $\left\vert L\right\rangle $, the scattering process is $%
\left\vert \phi (\pi /2,L)\right\rangle \left\vert L\right\rangle
\longrightarrow \left\vert \phi (-\pi /2,L)\right\rangle \left\vert
L\right\rangle $, i.e., the incoming wave packet is totally reflected and
the qubit remains to be in state $\left\vert L\right\rangle $. These two
processes are illustrated schematically in Fig. 3 (b, c). Remarkably, if a
sequent wave packets scatter with the BP qubit in $\left\vert R\right\rangle 
$\ state, the first one can pass freely, but the subsequent ones will be
reflected totally. This suggests a simple scheme to realize a quantum switch
to control the coherent transport of a photon in a one-dimensional
waveguide. The photon blockade \cite{Birnbaum} can be utilized to construct
a photon-pair qubit in coupled-cavity array. Nevertheless, different from
schemes in Refs. \cite{Chang, ZhouL}, our scheme allows one and \textit{only
one} photon passing over the switch.

%%%%%%%%%%%%%%%%%%%%%%%%%%%%%%%%%%%%%%%%%%%%%%%%%%%%%%%%%%%%%%%%%%%%%%%%
\begin{figure}[tbp]
\includegraphics[ bb=50 270 540 710, width=7.0 cm, clip]{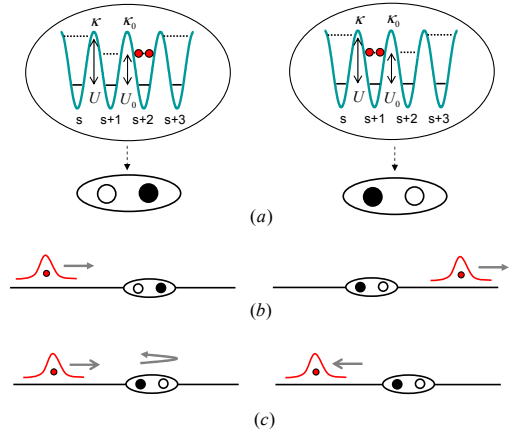}
\caption{\textit{(Color online) (a) A confined BP as a charge qubit with
states }$\left\vert R\right\rangle $\textit{\ and }$\left\vert
L\right\rangle $\textit{. (b, c) A BP qubit as a quantum switch to control
the transport of a SP wave packet. In the case with the qubit in state }$%
\left\vert R\right\rangle $\textit{, a moving SP wave packet will pass
through the qubit freely but leaves the qubit to be in the }$\left\vert
L\right\rangle $\textit{\ state. In contrast, if the qubit is in state }$%
\left\vert L\right\rangle $\textit{, the coming wave packet will be totally
reflected and remains the qubit to be in state }$\left\vert L\right\rangle $%
\textit{. }}
\label{fig3}
\end{figure}
%%%%%%%%%%%%%%%%%%%%%%%%%%%%%%%%%%%%%%%%%%%%%%%%%%%%%%%%%%%%%%%%%%%%%%%%%

\textit{Conclusion. }In conclusion, we have studied the existence of
localized BP in Bose Hubbard model with arbitrary on-site interaction $U$.
We have shown that BP wave packets refer to different regimes of a bound
pair band (BPB)\ and have different group velocities as $U$ varies from zero
to infinity, but spread to the same fidelity when they travel over the same
distance. It proposed a new object to be a flying or stationary qubit in
quantum device. Furthermore, the coherent shift in large-$U$ system suggests
a BP qubit as a quantum switch embeded in a one-dimensional waveguide. Our
analysis can be extended to a fermion Hubbard system with a minor correction.

We acknowledge the support of the CNSF (grant No. 10874091, 2006CB921205).


\begin{thebibliography}{\dag}
\bibitem[\dag]{email} emails: songtc@nankai.edu.cn \newline

\bibitem{Winkler} K. Winkler \textit{et al.}, Nature London \textbf{441},
853 (2006).

\bibitem{Mahajan} S. M. Mahajan and A. Thyagaraja, J. Phys. A \textbf{39},
L667 (2006).

\bibitem{Petrosyan} D. Petrosyan, \textit{et al.}, Phys. Rev. A \textbf{76},
033606 (2007).

\bibitem{Creffield} C. E. Creffield, Phys. Rev. A \textbf{75}, 031607(R)
(2007).

\bibitem{Kuklov} A. Kuklov and H. Moritz, Phys. Rev. A \textbf{75}, 013616
(2007).

\bibitem{Folling} S. F\"{o}lling \textit{et al.}, Nature \textbf{448}, 1029
(2007).

\bibitem{Zollner} S. Z\"{o}llner \textit{et al.}, Phys. Rev. Lett. \textbf{%
100}, 040401 (2008).

\bibitem{ChenS} L. Wang \textit{et al.}, Eur. Phys. J. D \textbf{48}, 229
(2008).

\bibitem{YangCN} C. N. Yang, Phys. Rev. Lett. \textbf{63}, 2144 (1989).

\bibitem{Jaksch} Jaksch \textit{et al}. Phys. Rev. Lett. \textbf{81}, 3108
(1998).

\bibitem{QBEC} M. Greiner \textit{et al.}, Nature \textbf{415}, 39 (2002).

\bibitem{cavities} M.J. Hartmann, \textit{et al.}, Nature Phys. \textbf{2},
849 (2006); M.J. Hartmann \textit{et al.}, Phys. Rev. Lett. \textbf{99},
103601 (2007).

\bibitem{LY} Y. Li, \textit{et al.}, Comm. Theor. Phys. \textbf{48}, 445
(2007).

\bibitem{YS} S. Yang, \textit{et al.}, Phys. Rev. A \textbf{71}, 022317
(2006).

\bibitem{Birnbaum} K.M. Birnbaum, \textit{et al.,} Nature \textbf{436}, 87
(2005).

\bibitem{Chang} D. E. Chang \textit{et al.}, Nature Phys. \textbf{3}, 807
(2007).

\bibitem{ZhouL} L. Zhou, \textit{et al.}, Phys. Rev. Lett. \textbf{101},
100501 (2008).
\end{thebibliography}
\end{document}